# multiPI-TransBTS: A Multi-Path Learning Framework for Brain Tumor Image Segmentation Based on Multi-Physical Information

Hongjun Zhu[a,b,c]*, Jiaohang Huang[a,b], Kuo Chen[a,b], Xuehui Ying[a,b], Ying Qian[a,b]

[a] School of Software Engineering, Chongqing University of Posts and Telecommunications, Chongqing 400065, China

[b] Chongqing Engineering Research Center of Software Quality Assurance, Testing and Assessment, Chongqing 400065, China

[c] Key Laboratory of Big Data Intelligent Computing, Chongqing University of Posts and Telecommunications, Chongqing 400065, China

E-mail: zhuhj@cqupt.edu.cn

*Corresponding author: Hongjun Zhu.

**Abstract**

Brain Tumor Segmentation (BraTS) plays a critical role in clinical diagnosis, treatment planning, and monitoring the progression of brain tumors. However, due to the variability in tumor appearance, size, and intensity across different MRI modalities, automated segmentation remains a challenging task. In this study, we propose a novel Transformer-based framework, multiPI-TransBTS, which integrates multi-physical information to enhance segmentation accuracy. The model leverages spatial information, semantic information, and multi-modal imaging data, addressing the inherent heterogeneity in brain tumor characteristics.

The multiPI-TransBTS framework consists of an encoder, an Adaptive Feature Fusion (AFF) module, and a multi-source, multi-scale feature decoder. The encoder incorporates a multi-branch architecture to separately extract modality-specific features from different MRI sequences. The AFF module fuses information from multiple sources using channel-wise and element-wise attention, ensuring effective feature recalibration. The decoder combines both common and task-specific features through a Task-Specific Feature Introduction (TSFI) strategy, producing accurate segmentation outputs for Whole Tumor (WT), Tumor Core (TC), and Enhancing Tumor (ET) regions.

Comprehensive evaluations on the BraTS2019 and BraTS2020 datasets demonstrate the superiority of multiPI-TransBTS over the state-of-the-art methods. The model consistently achieves better Dice coefficients, Hausdorff distances, and Sensitivity scores, highlighting its effectiveness in addressing the BraTS challenges. Our results also indicate the need for further exploration of the balance between precision and recall in the ET segmentation task. The proposed framework represents a significant advancement in BraTS, with potential implications for improving clinical outcomes for brain tumor patients.

# 1. INTRODUCTION

Brain tumors, although relatively rare compared to other cancers, pose significant clinical challenges due to their complex and often aggressive nature. Magnetic resonance imaging (MRI) is the only test needed to diagnose a brain tumor, as certain low-grade tumors, such as astrocytomas, not visible on computer tomography (CT) scans, may be detected by MRI [1]. MRI allows for tumor volume measurement and assessment of the tumor's relationship with critical brain structures, including blood vessels [2]. Consequently, MRI is regarded as the gold standard due to its high spatial resolution and contrast differentiation [3].

These MRI scans include pre- and post-contrast T1-weighted (T1 and T1Gd), T2-weighted (T2), and T2 Fluid Attenuated Inversion Recovery (T2-FLAIR) volumes [4]. T1Gd is acquired with intravenous gadolinium contrast, and enhancing regions indicate disruption (or absence) of the blood-brain barrier, which is consistent with viable tumor tissue and infiltrated brain [5]. Gadolinium enhancement in post-contrast T1-weighted images reveals focal areas where the blood-brain barrier is compromised, although it may not reveal larger areas of infiltrating tumor [6, 7]. Furthermore, some high-grade gliomas show no gadolinium enhancement [7, 8]. Since different MRI modalities capture distinct characteristics of the underlying anatomy, clinical analysis typically combines multiple MRI modalities for diagnosis and treatment planning [9, 10].

It is widely accepted that a higher intensity of enhancement, larger areas of necrosis, and edema are associated with higher-grade gliomas and poorer prognoses [5]. Therefore, Brain Tumor Segmentation (BraTS) of MRI scans is essential for clinical diagnosis, treatment planning, and disease monitoring [11, 12]. Accurate tumor delineation aids in planning surgical resection, radiotherapy, and other treatments, enabling surgeons to maximize tumor removal while minimizing damage to healthy brain tissue. Additionally, segmentation allows for the assessment of tumor progression or regression over time, facilitating treatment evaluation and adjustment of therapeutic strategies.

Despite its importance, BraTS is an challenging task. Brain tumors vary widely, with over 20 different types [2]. The tumor variability in appearance, size, location, and intensity, coupled with the similarity between tumor and non-tumor tissue in imaging, presents significant challenges [11]. Furthermore, the difficulty in estimating precise tumor boundaries during surgery is reflected in segmentation labels, resulting in high uncertainty among experts in delineating these boundaries [13]. Manual segmentation is often time-consuming and subjective, emphasizing the need for automated brain tumor segmentation methods [14].

To address this, researchers have approached BraTS from various angles. Existing methods focus on aspects such as intensity, gradient, shape, contour, texture, and symmetry [3]. These methods range from threshold-based, feature-based, contour-based, and region-based to learning-based approaches, with learning-based methods proving the most effective for BraTS [3].

However, current learning-based methods do not fully leverage available information. Specifically: (1) different brain tumor regions exhibit distinct characteristics across different MRI sequences, and (2) different MRI modalities vary in their performance across tumor region segmentation tasks. This limitation impedes further improvements in segmentation accuracy, particularly for boundary voxels that lack contextual information [14].

In response to these challenges, we propose multiPI-TransBTS, an integrative framework tailored for the BraTS task. This model incorporates multi-physical MRI information within a multi-

task learning framework. Our main contributions include:

(1) We propose a Transformer-based framework that integrates multi-physical information for BraTS, reducing uncertainty in model representation and thereby improving segmentation accuracy.

(2) We construct a multi-branch network architecture to extract modality-specific features separately, avoiding interference from irrelevant modalities in specific BraTS tasks.

(3) We design an Adaptive Feature Fusion (AFF) module to fuse information from different MRI modalities, forming multi-scale features shared across tasks.

(4) We develop a multi-source and multi-scale feature decoder, which respects the differences between segmentation tasks and fully utilizes both common and individual features.

(5) We conduct comprehensive evaluations using real-world datasets. Our experiments on the BraTS2019 and BraTS2020 datasets demonstrate multiPI-TransBTS's superior performance over existing methods in terms of Dice coefficient, Hausdorff distance, and Sensitivity. To facilitate further research, the source code for the multiPI-TransBTS framework is available at https://github.com/JoetheReindeer/multiPI_TransBTS.

# 2 RELATED WORK

MRI is a widely used imaging technique to assess these tumors, but the large amount of data produced by MRI prevents manual segmentation in a reasonable time, limiting the use of precise quantitative measurements in clinical practice [15]. Manual segmentation of brain tumor extent from 3D MRI volumes is a very time-consuming task and the performance is highly reliant on the operator's experience [16]. In this context, a reliable fully automatic segmentation method for brain tumor segmentation is necessary for an efficient measurement of the tumor extent[16]. For this reason, many research works have been undertaken to apply deep learning techniques for the BraTS task. These techniques can be classified roughly into four categories: CNN-based methods, RNN-based methods, GAN-based methods, and Transformer-based methods.

## 2.1 CNN-based methods

Convolutional Neural Networks (CNNs) inherently incorporate the spatial hierarchy of features within an image. Utilizing local receptive fields and weight sharing, CNNs embed the prior knowledge that similar patterns, such as edges or textures, are likely to recur across various parts of a brain tumor. This underpinning principle has facilitated the expansion of CNN applications within the BraTS domain, leading to the development of numerous models tailored to address the intricacies of these tasks.

## 2.1.1 2D CNN

Traditional CNNs generally consist of several convolutional layers, followed by fully connected layers at the end to output a single label. To adapt this architecture for direct image-image mapping, Fully Convolutional Neural Networks (FCNNs) replace fully connected layers with additional convolutional layers, enhancing their utility in the BraTS challenges. For example,

Pereira et al. [15] proposed an automatic segmentation method using CNNs that leverage small 3 × 3 kernels, enabling deeper network architectures while mitigating overfitting due to the reduced number of weights. Kamnitsas et al. [17] introduced a dual pathway architecture that processes images at multiple scales to incorporate both local and contextual information more effectively. Zhao et al. [18] combined FCNNs with Conditional Random Fields (CRFs) to achieve segmentation results with appearance and spatial consistency.

Li et al. [19] developed an FCNN based on the U-Net architecture augmented with inception modules, optimizing the model for asymmetrical tumor regions. To further enhance model accuracy, Chen et al. [20] incorporated a Left-Right Similarity Mask (LRSM) into their FCNNs, addressing the inherent asymmetry in tumor imaging. Zhou et al. [14] utilized multi-task networks to distribute common information effectively and address class imbalance within the data. More recent developments include Cinar et al.'s DenseNet-UNet hybrid model [21] and Ullah et al.'s Multiscale Residual Attention-UNet (MRA-UNet) [22], both designed to refine BraTS segmentation accuracy. In addition, Allah et al. [23] introduced the U-Net model for enhanced localization of tumors, and Rehman et al. [24] proposed the RAAGR2-Net with a Residual Spatial Pyramid Pooling (RASPP) module to preserve location information across network layers.

## 2.1.2 3D CNN

3D CNNs offer a solution to the slice-level inconsistencies resulting from 2D CNNs by harnessing the three-dimensional continuity of MRI data. Chen et al. [25] proposed the Multi-Level DeepMedic model that utilizes multi-level information to achieve more precise segmentation. Isensee et al. [26] developed nnU-Net, an adaptable and self-configuring system designed to automatically adjust to various medical imaging tasks without manual intervention. This model effectively addresses the diverse challenges presented by different medical imaging datasets.

Additionally, Li et al. [27] suggested the use of cascaded 3D U-Nets for enhanced performance in BraTS tasks, while Chang et al. [28] designed a residual dual-path attention-fusion 3D CNN to amalgamate global and local channel information. Raza et al. [29] introduced the dResU-Net, combining features of residual networks and U-Net for robust segmentation capabilities.

Despite their potential, 3D CNNs are often constrained by their substantial computational demands and the network size required, which can become prohibitive, particularly with anisotropic datasets [26]. Therefore, 2D CNNs continue to be a popular choice due to their reduced computational requirements and robust performance across varying imaging conditions [18].

In summary, CNNs, through their architectural design, introduce general priors concerning spatial hierarchies, translation invariance, and local feature consistency. U-Net, in particular, brings additional specific priors about the importance of multi-scale features and the integration of detailed and contextual information within its unique architecture, proving essential for complex segmentation tasks like those found in BraTS.

## 2.2 GAN-based methods

Generative Adversarial Networks (GANs) have significantly enhanced the BraTS performance by generating synthetic images that closely resemble authentic ones, thereby expanding the training dataset and reducing overfitting [30]. This advancement facilitates a wide application in the BraTS field.

Li et al. [31] introduced TumorGAN, a framework for generating image segmentation pairs based on unpaired adversarial training. This method allows for the creation of accurate and diverse training images from limited datasets. Zhu et al. [9] advanced this approach by developing a dual-scale GAN capable of generating multi-modality images, thus enriching the training data further and allowing models to learn from a broader range of image types. Similarly, Jia et al. [12] utilized GANs to produce super-resolution images, enhancing the detail available in MRI scans which is crucial for identifying subtle features of brain tumors. Hamghalam and Simpson [32] took a different approach by using a conditional GAN to specifically enhance the contrast of tumor subregions in MRI scans. This technique conditions the GAN on additional target information such as segmentation maps, improving the quality and utility of generated images for training segmentation models.

On the other hand, Xue et al. [33] proposed an adversarial critic network designed to capture both global and local features of the images, focusing on long- and short-range spatial relationships between pixels. This approach enhances the GAN's ability to produce more detailed and clinically relevant images. Nema et al. [34] designed an architecture named RescueNet, which segments the whole tumor and its subregions such as the core and enhanced areas, providing a more comprehensive tool for tumor analysis in brain MRI scans. Additionally, Ding et al. [35] introduced a two-stage generator to refine the brain tumor segmentation performance further. Cui et al. [36] developed a generator based on an encoder-decoder structure, which simultaneously generates segmentation maps and reconstructs original images.

In summary, GANs represent a powerful tool for data augmentation and improving segmentation quality. However, despite their significant potential, GANs are also known for their training instability, which can lead to inconsistent quality in the segmentation results. This remains an area of active research.

## 2.3 RNN-based methods

Recurrent Neural Networks (RNNs) are initially designed to handle sequential data, which have the same properties as MRI slices. RNN can introduce the prior knowledge that the input data has a temporal or sequential relationship. This is particularly relevant for the BraTS task, where consecutive slices of the brain may show the gradual growth or movement of a tumor.

To harness this sequential data effectively, Deng et al. [37] integrated a Conditional Random Field with a Recurrent Neural Network (CRF-RNN). This combination leverages the sequential dependencies across slices to improve the consistency and accuracy of segmentation.

Long Short-Term Memory (LSTM) networks, an advanced form of RNNs, are specifically designed to handle long-term dependencies within sequential data. This ability is crucial in brain tumor segmentation, where characteristics of distant slices may influence the interpretation and segmentation of subsequent slices.

Building on these capabilities, Hu et al. [38] proposed the UNET-LSTM algorithm, which aims to address the challenge of sample imbalance in the dataset. By integrating LSTM with the U-Net architecture, this approach enhances the model's ability to predict more balanced and accurate segmentations across the dataset.

In summary, RNN and LSTM models are particularly valuable in brain tumor segmentation due to their ability to capture and utilize the sequential and contextual information inherent in MRI slice series. These models presuppose a significant continuity in the features and progression, allowing for a more nuanced understanding and representation of tumor evolution between consecutive slices.

## 2.4 Transformer-based methods

Unlike CNNs, which primarily focus on local correlations through convolutions, Transformers excel in capturing global context due to their self-attention mechanism. This mechanism can model relationships between distant regions in an input image, providing a comprehensive understanding of spatial contexts. Unlike RNNs, which process data sequentially, Transformers can handle different parts of the image in parallel, effectively managing long-range dependencies.

To fully exploit the merits of both Transformers and CNNs, numerous Transformer-CNN hybrid models have been developed. These models combine the global contextual capabilities of Transformers with the robust local feature extraction of CNNs, particularly leveraging the U-shaped architecture.

For instance, TransUNet [39] integrates the self-attention mechanism of Transformers with the encoding-decoding structure of UNet. This was one of the first models to effectively blend local and global information for segmentation accuracy increase. UNETR [40] employs a stack of transformers as the encoder, connected to a decoder via skip connections. This design allows for an effective synthesis of multi-level feature information. CKD-TransBTS [5] features a dual-branch hybrid encoder and a feature calibration decoder within a U-Net-like structure, integrating features at various scales.

In addition, TranSiam [41] consists of two identical sub-networks where convolutions extract detailed information at lower levels, and Transformers handle global information processing at higher levels. SDV-TUNet [42] utilizes multi-head self-attention and sparse dynamic adaptive fusion to meticulously extract global spatial semantic features, crucial for precise BraTS.

Traditional Transformers operate on fixed-size patches, which can somewhat restrict their ability to process multi-scale information. To overcome this limitation, Swin Transformer utilizes a hierarchical feature representation strategy to capture both local and global information adeptly [43]. Similarly, IMS2Trans [44] employs Swin Transformer technology to enable efficient information sharing and fusion among different modalities.

In summary, the application of Transformers in brain tumor segmentation leverages their ability to capture global context and handle multi-scale information while maintaining minimal inductive biases. The self-attention mechanism allows these models to dynamically focus on crucial regions of the image, making them exceptionally suitable for the complex BraTS task. However, current methods predominantly focus on exploiting the generic capabilities of models and overlook the integration of domain-specific knowledge related to brain tumor imaging. This oversight limits further enhancements in the BraTS performance, suggesting a need for more specialized approaches that incorporate specific clinical and imaging insights into the model architecture.

# 3 MRI PRINCIPLES OF BRAIN TUMORS

## 3.1 Physical principles of MRI

MRI exploits the principles of nuclear magnetic resonance to generate detailed images of the body. In MRI, atomic nuclei with odd numbers of protons or neutrons, such as hydrogen, possess nuclear spin with a quantum mechanical property. By applying a radiofrequency pulse at a specific frequency, which corresponds to the energy difference between two states, these nuclei can be excited from their lower energy state to a higher one. This phenomenon is known as nuclear magnetic resonance [45].

Upon removal of the RF pulse, the nuclei return to their lower energy state through a process called relaxation, which occurs in two primary forms: spin-lattice relaxation and spin-spin relaxation. Spin-lattice relaxation releases absorbed energy to the surrounding molecular lattice, returning the nuclei to thermal equilibrium. The rate of this relaxation is measured by the T1 relaxation time. Spin-spin relaxation involves the dephasing of spins in the transverse plane due to interactions among the spins themselves, without energy transfer to the lattice. The rate of spin-spin relaxation is characterized by the T2 relaxation time [45].

T1-weighted imaging and T2-weighted imaging are techniques used to highlight different tissue characteristics based on their T1 and T2 relaxation times. T1-weighted imaging employs short repetition time (TR) and short echo time (TE) to emphasize differences in T1 properties between different types of tissue. T2-weighted imaging utilizes long TR and long TE to accentuate variations in T2 relaxation time [45].

To enhance the diagnostic capabilities of MRI, other two techniques tend to be used at the same time. One is gadolinium contrast-enhanced T1-weighted imaging (T1Gd), in which Gadolinium enhances the contrast by shortening the T1 relaxation time of nearby water protons, making the affected areas appear brighter in the images [46]. The other one is T2 Fluid Attenuated Inversion Recovery (FLAIR). This technique involves an inversion recovery pulse that nulls the signal from fluids, particularly cerebrospinal fluid, to suppress the background fluid signal and enhance the detection of lesions[47].

Each of these MRI techniques provides a unique perspective on tissue properties and pathological changes. Moreover, tumors' borders are often fuzzy and hard to distinguish from healthy tissue [48]. For these reasons, medical analysis and diagnosis are usually carried out in combination with multiple MRI modalities [9, 10].

## 3.2 Structure of brain tumors

Brain tumors refer to a diverse collection of intracranial neoplasms, comprising over 20 distinct types, each with its unique biology [1]. The MRI images of brain tumors, illustrated in Fig. 1, are composed of four primary sub-regions: necrotic core (NCR, green), edema (ED, yellow), non-enhancing tumor core (NET, red), and GD-enhancing tumor core (ET, blue) [4, 49]. Each of these sub-regions plays a crucial role in the clinical diagnosis and treatment of brain tumors [50]. It is important to note that these MRI sub-regions do not strictly represent biological entities, but are rather image-based constructs. Often, there is limited evidence in the imaging data for the presence of the non-enhancing solid core [13]. For this reason, the non-enhancing solid core has been excluded from the 2017-present BraTS dataset.

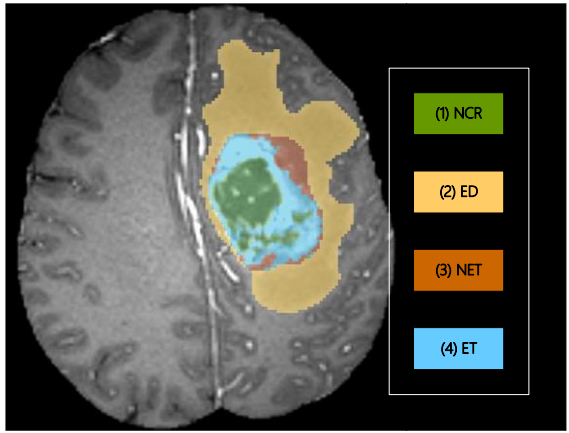

Fig. 1. The structure of brain tumor. The MRI images of a brain tumor consist of four distinct sub-regions: (1) necrotic core (NCR, green), (2) edema (ED, yellow), (3) non-enhancing tumor core (NET, red), and (4) GD-enhancing tumor core (ET, blue), also referred to as the active tumor core. Figure adapted from [4].

Driven by the need for clinical relevance, the Multimodal Brain Tumor Image Segmentation Benchmark (BRATS), a well-established community benchmark, focuses on the segmentation of three tumor regions: (1) the whole tumor (WT), which includes all four sub-regions; (2) the tumor core (TC), which excludes edema but includes the other three sub-regions; and (3) the GD-enhancing tumor core (ET) [13]. The relationship between the segmented regions and the annotated sub-regions in the BRATS datasets is summarized in Table 1.

Table 1. The relationship between the segmented regions and the annotated sub-regions in the BRATS datasets

|    | NCR/NET(label 1) | ED (label 2) | ET(label 4) |
|----|------------------|--------------|-------------|
| WT | 1                | 1            | 1           |
| TC | 1                | 0            | 1           |
| ET | 0                | 0            | 1           |

Different MRI modalities exhibit varying sensitivities to different types of tissue, suggesting that their contributions may differ in segmenting various tumor regions. To investigate this, we evaluated the impact of different MRI modalities on the segmentation performance for the three tumor-region categories using the classic U-Net model [51]. The results obtained from the BRATS dataset, shown in Fig. 2, are consistent with findings from Yang [52], indicating that different modalities perform differently across the tumor region segmentation tasks.

# 4 METHODOLOGY

Building on the observation that different modalities exhibit variable performance across tumor region segmentation tasks, we propose multiPI-TransBTS, a Transformer-based framework that integrates multi-physical information for the BraTS task. The framework leverages common image data attributes such as spatial and semantic information, alongside specific information relevant to BraTS, like multimodal imaging knowledge.

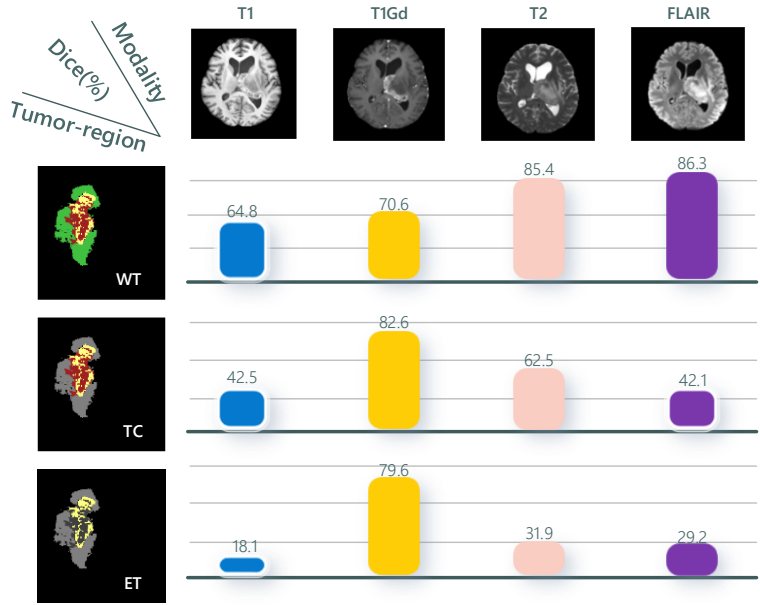

Fig. 2. Comparison of segmentation performance of three tumor-region categories using different modalities. WT, TC, and ET represent the whole tumor, tumor core, and active tumor, respectively. In the segmentation maps, green corresponds to the edema, yellow represents the enhancing tumor, and red indicates the non-enhancing tumor and necrosis.

## 4.1 Overall framework

The overall model design is based on several key considerations: (1) CNNs are effective at capturing spatial information and local patterns, so multiPI-TransBTS predominantly utilizes CNN architectures; (2) As image data are high-dimensional and often contain redundant information, such as similar pixel values in smooth regions, multiPI-TransBTS adopts an encoding-decoding structure; (3) Given the heterogeneity in tumor size and shape, multiPI-TransBTS incorporates skip connections from the U-Net architecture to handle the significant variation in tumor characteristics; (4) Since different brain tumor regions exhibit distinct characteristics across different MRI scan modalities, multiPI-TransBTS employs a multi-branch network architecture to extract modality-specific features separately; and (5) recognizing that different modalities exhibit varying performance across tumor region segmentation tasks, multiPI-TransBTS implements a Task-Specific Feature Introduction (TSFI) strategy.

The overall framework of multiPI-TransBTS, shown in Fig. 3, is divided into three main components: the encoder, the fusion module, and the decoder.

**I. Transformer-based encoder**

The encoder is responsible for capturing context and extracting features from the input image. Early encoder layers focus on fine details, while deeper layers capture more global context. To simplify feature expression, features are divided into common features (shared across tasks) and individual features (specific to each task), represented by the red and blue F-marked blocks in Fig. 3.

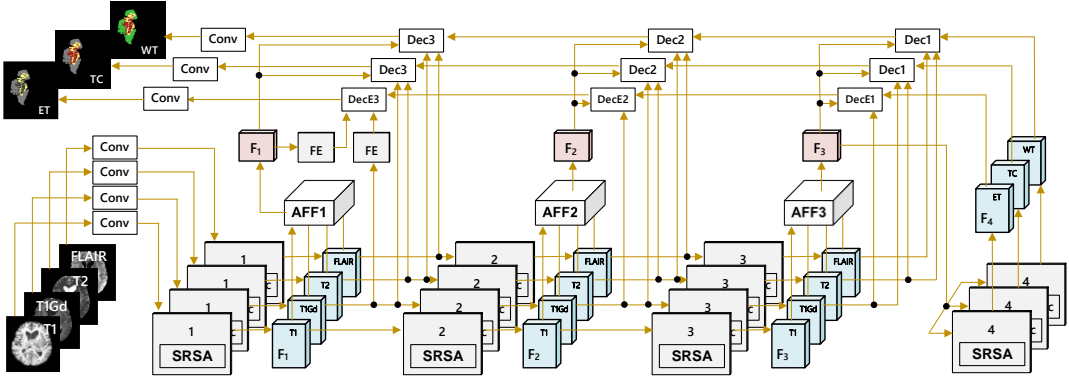

Fig. 3. Framework of multiPI-TransBTS. This framework consists of three main modules: an encoder, a fusion module, and a decoder. To simplify feature representation, features are divided into common features shared across tasks and individual features specific to each task, denoted by red and blue F-marked blocks, respectively. "SRSA" indicates the Spatial-Reduction Self-Attention module. "AFF" and "FE" stand for adaptive feature fusion and feature enhancement. "Conv" and "Dec" are abbreviations for convolution and decoder, respectively.

The traditional U-Net architecture consists of convolutional layers followed by pooling layers, which capture features at multiple scales [51]. To allow the model to focus more on relevant parts of the input, multiPI-TransBTS incorporates self-attention mechanisms. Though the Vision Transformer (ViT) [53] has successfully introduced self-attention mechanisms into image classification tasks, its tendency to produce low-resolution outputs makes it unsuitable for dense prediction tasks.

To overcome this, we use the Spatial-Reduction Self-Attention (SRSA) modules from the Pyramid Vision Transformer (PVT) [54], which can process dense partitions of an image and achieve high output resolution through a progressively shrinking pyramid structure. However, applying PVT directly to the BraTS task is impractical because PVT was initially designed for single-image segmentation and lacks the capability to fuse multimodal information required in BraTS. Therefore, multiPI-TransBTS leverages SRSA modules to reconstruct the encoder architecture, ensuring effective multimodal information integration.

### II. Adaptive feature fusion module

The fusion module in multiPI-TransBTS integrates complementary information from multiple sources or modalities into a unified representation. It performs multi-modality fusion across multi-scale features extracted by the backbone network, allowing fusion at different resolutions or layers. This strategy effectively combines fine-grained and coarse information, enabling the model to capture a broad range of contextual details.

To enhance feature representation, the fusion module uses Squeeze-and-Excitation (SE) mechanisms [55], allowing the network to recalibrate channel-wise importance and emphasize key features while suppressing less informative ones. Additionally, it incorporates an element-wise attention mechanism to amplify crucial spatial information, leading to a more accurate representation of important details in the input data.

This combination of channel-wise and element-wise attention mechanisms strengthens the model's ability to focus on critical features across both spatial and channel dimensions, thereby improving performance in segmentation tasks.

### III. Multi-feature decoder

The decoder integrates both common and individual features from various tasks to reconstruct the abstract feature representation into a segmentation output that matches the dimensionality of the input.

To retain fine details and high-resolution information, the decoder uses skip connections to directly link encoder layers to corresponding decoder layers. Furthermore, feature maps are upsampled before applying standard convolution, reducing the risk of checkerboard artifacts, as the upsampling process is separate from the convolution.

multiPI-TransBTS effectively introduces multi-physical information through tailored architectures, multi-level attention, and multimodal fusion, achieving accurate and efficient BraTS segmentation.

The core modules in Fig. 3, such as "SRSA", "AFF", and decoder, will be discussed in more detail in sections 4.2, 4.3, and 4.4, respectively.

## 4.2 Transformer-based encoder

The encoder in multiPI-TransBTS consists of an initial convolutional layer followed by a PVT-like architecture. This architecture is organized into four stages, each responsible for generating feature maps at different scales, as illustrated in Fig.3. To effectively capture the features from each modality separately, the first three stages include four branches, while the final stage is an exception. Each branch follows a similar structure, represented by the "SRSA" module in Fig. 3, which includes an Embedding Layer and Transformer Encoder. The internal details of the "SRSA" module are depicted in Fig. 4.

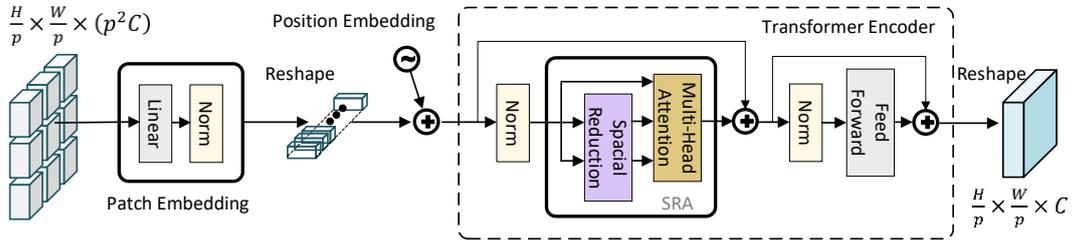

Fig. 4. "SRSA" module in Pyramid Vision Transformer (PVT). This module represents the Spatial-Reduction Self-Attention (SRSA) mechanism, which is responsible for reducing the spatial dimension of input sequences while capturing relevant attention.

### I. Embedding Layer

Given input scans **X** with size $H \times W \times C$, where $H$, $W$, and $C$ represent the height, width, and number of channels, respectively, the image is first divided into $\frac{HW}{p^2}$ patches, each of size $p \times p \times C$. Then, a patch embedding operation is performed, which can be formalized as:

$$\boldsymbol{\alpha} = \text{Norm}(\mathbf{X}\mathbf{W}^{\text{T}} + \mathbf{b}), \tag{1}$$

where Norm denotes Layer Normalization, and **W**, **b** are learnable parameters. The resulting **α** is reshaped, and positional embeddings are added before being input to the Transformer encoder. These operations can be formulated as:

$$\boldsymbol{\beta} = \text{Reshape}(\boldsymbol{\alpha}) + \mathbf{P}. \tag{2}$$

Here, Reshape(.) is the reshaping operation, and **P** represents the positional embeddings, which are element-wise added to the input features to inject positional information into the model. These embeddings are initialized as trainable parameters and learned during the training process.

### II. Transformer Encoder

The Transformer encoder consists of two residual networks. The first residual network can be expressed as:

$$\gamma = \text{SRA}(\text{Norm}(\boldsymbol{\beta})) + \boldsymbol{\beta}, \tag{3}$$

where SRA indicates spatial-reduction attention, formulated as:

$$\text{SRA}(\text{Norm}(\boldsymbol{\beta})) = \text{Concat}(\mathbf{h}_1, \mathbf{h}_2, \cdots, \mathbf{h}_k)\mathbf{W}^O \tag{4}$$

with $\mathbf{h}_i$ given by:

$$\mathbf{h}_i = \text{Softmax}\left(\frac{\left((\text{Norm}(\boldsymbol{\beta}))\mathbf{W}_i^Q\right)\left(\text{SR}(\text{Norm}(\boldsymbol{\beta}))\mathbf{W}_i^K\right)^T}{\sqrt{d_k}}\right)\left(\text{SR}(\text{Norm}(\boldsymbol{\beta}))\mathbf{W}_i^V\right). \tag{5}$$

Here, $\text{Concat}(\cdot)$ denotes the concatenation operation, and $\text{SR}(\cdot)$ is the operation for reducing the spatial dimension of the input sequence, expressed as:

$$\text{SR}(\text{Norm}(\boldsymbol{\beta})) = \text{Norm}(\text{Reshape}(\text{Norm}(\boldsymbol{\beta}))\mathbf{W}^S), \tag{6}$$

where $\mathbf{W}^O$, $\mathbf{W}_i^Q$, $\mathbf{W}_i^K$, $\mathbf{W}_i^V$, and $\mathbf{W}^S$ are learnable weight matrices.

The second residual network is defined as:

$$\boldsymbol{\rho} = \text{FFN}(\text{Norm}(\boldsymbol{\gamma})) + \boldsymbol{\gamma}, \tag{7}$$

where $\text{FFN}(.)$ denotes the Feed-Forward Network, given by:

$$\text{FFN}(\text{Norm}(\boldsymbol{\gamma})) = \text{Max}(0; (\text{Norm}(\boldsymbol{\gamma}))\mathbf{W}_A + \mathbf{b}_A)\mathbf{W}_B + \mathbf{b}_B + \boldsymbol{\gamma}. \tag{8}$$

Here, the parameters $\mathbf{W}_A$, $\mathbf{W}_B$, $\mathbf{b}_A$, and $\mathbf{b}_B$ are learnable weight matrices and bias terms.

Finally, the output $\mathbf{Y}$ of the Transformer Encoder is given by:

$$\mathbf{Y} = \text{Reshape}(\boldsymbol{\rho}) \tag{9}$$

This patch embedding operation is applied across multiple stages, progressively shrinking the scale of feature maps. By adjusting the scale of the feature map in each stage, the encoder constructs a feature pyramid, enhancing its ability to handle features at different resolutions.

## 4.3 Adaptive feature fusion module

The Adaptive Feature Fusion (AFF) module is designed to fuse information from different modalities of MRI scans, forming multi-scale features shared across multiple tasks. As shown in

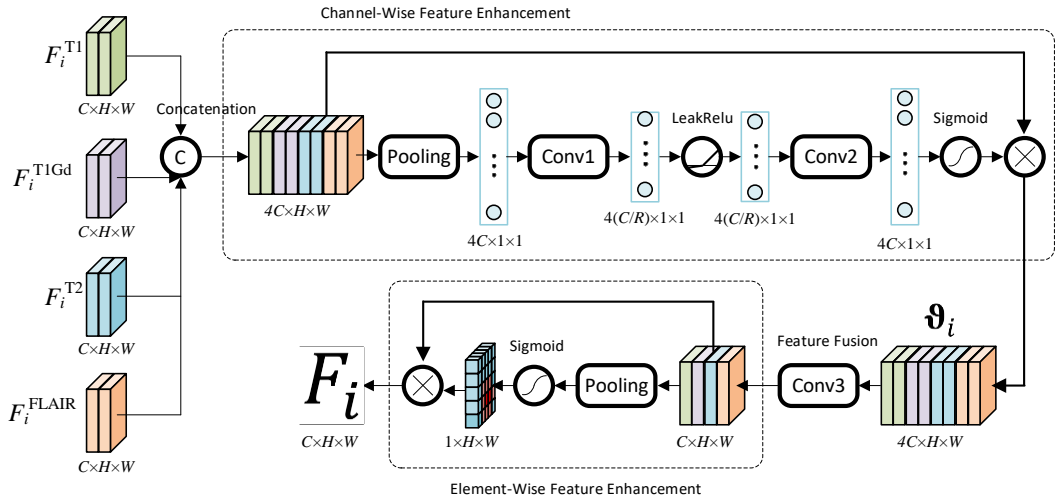

Fig. 5. Adaptive Feature Fusion (AFF) Module. This module integrates complementary information from different modalities. It combines channel-wise and element-wise attention to enhance feature representation.

Fig. 5, the AFF module consists of three main submodules: channel-wise feature enhancement, feature fusion, and element-wise feature enhancement. The channel-wise feature enhancement uses Squeeze-and-Excitation (SE) Networks, the feature fusion is performed using convolutional networks, and the element-wise feature enhancement employs feature recalibration.

**I. Channel-Wise Feature Enhancement**

Channel-wise feature enhancement is performed by the standard SE block. In this block, the input feature maps from different modalities are concatenated to combine the information from multiple modalities. Given the feature maps from T1, T1Gd, T2, and FLAIR generated by the "SRSA" module at the $i$-th stage, represented as $\mathbf{F}_i^{T1}, \mathbf{F}_i^{T1Gd}, \mathbf{F}_i^{T2}, \mathbf{F}_i^{FLAIR}$, see the blue blocks in Fig. 3. This operation can be formulated as:

$$\boldsymbol{\varphi}_i = \text{Concat}\big(\mathbf{F}_i^{T1}, \mathbf{F}_i^{T1Gd}, \mathbf{F}_i^{T2}, \mathbf{F}_i^{FLAIR}\big) \qquad (i = 1,2,3). \qquad (10)$$

Next, a global average pooling operation is applied to reduce the spatial dimensions from $H \times W$ to $1 \times 1$, summarizing the global information of each feature map. The pooled result is then passed through two convolutional layers. The first convolution reduces the channel dimensionality by a factor of $R$, followed by a non-linearity operation (Leaky ReLU). A second convolution layer restores the original dimensionality. The output is then passed through a Sigmoid function to generate channel-wise attention weights:

$$\boldsymbol{\theta}_i = \text{Sigmoid}\big(\text{Conv}_2\big(\text{LeakyReLU}\big(\text{Conv}_1(\text{Pooling}(\boldsymbol{\varphi}_i))\big)\big)\big) \qquad (i = 1,2,3). \qquad (11)$$

Here, Conv() denotes a convolutional operation, and Pooling() represents the pooling operation. LeakyReLU() and Sigmoid() are activation functions, where the former introduces non-linearity and the latter clamps the output between 0 and 1.

The generated weights $\boldsymbol{\theta}_i$ are used to recalibrate the original input feature maps, emphasizing the important features. After performing elementwise multiplication, we get

$$\boldsymbol{\vartheta}_i = \boldsymbol{\theta}_i \otimes \boldsymbol{\varphi}_i \qquad (i = 1,2,3). \qquad (12)$$

Here, $\otimes$ denotes elementwise multiplication.

**II. Feature Fusion**

Feature fusion is performed using a convolution operation to combine the multiple features. This can be mathematically expressed as:

$$\boldsymbol{\phi}_i = \text{Conv}_3(\boldsymbol{\vartheta}_i) \qquad (i = 1,2,3). \qquad (13)$$

**III. Element-Wise Feature Enhancement**

The element-wise feature enhancement recalibrates the input by multiplying it with weights generated through an average pooling operation, rather than the more complex SE operation. This process can be expressed as:

$$\mathbf{F}_i = \boldsymbol{\phi}_i \otimes \text{Sigmoid}(\text{Pooling}(\boldsymbol{\phi}_i)) \qquad (i = 1,2,3). \qquad (14)$$

## 4.4 Multi-feature decoder

The decoder in multiPI-TransBTS is responsible for fusing information from different scales and combining both common and individual features to produce the final BraTS results. The decoder consists of three decoding stages, with each stage including three fundamental operations: elementwise multiplication, concatenation, and convolution. These operations are combined in different ways depending on the specific segmentation task. The inputs to the modules also vary across hierarchical levels. This approach is referred to as the Personalized Feature Introduction (PCI) strategy. Fig. 6(a) illustrates the decoder for WT and TC segmentation, while Fig. 6(b) shows the decoder for ET segmentation.

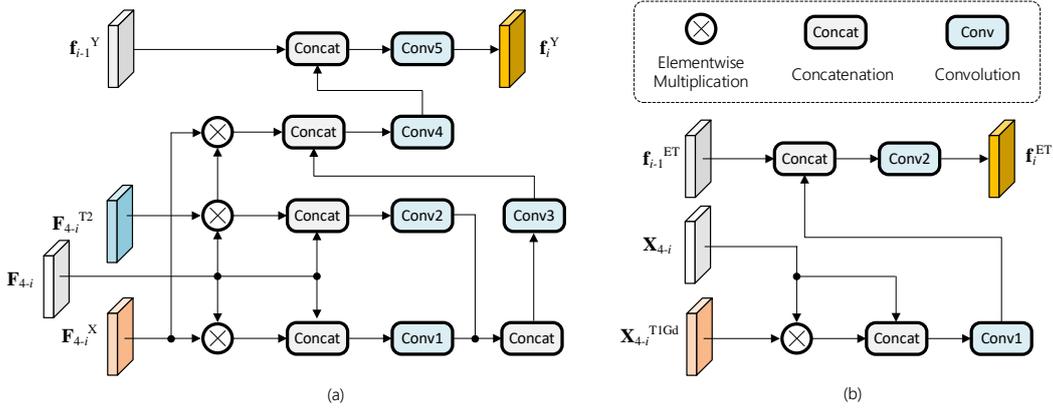

Fig. 6. Multi-feature decoder. (a) Decoder for WT and TC segmentation. (b) Decoder for ET segmentation.

**I. Decoders for WT and TC segmentation**

In multiPI-TransBTS, a multi-level decoder is used to achieve multi-scale information fusion. The decoders for WT and TC segmentation, see the "Dec" module in Fig. 3, have similar internal structures, as depicted in Fig. 6(a). Based on the correlation analysis between region segmentation and the MRI modalities (see Fig. 2), the WT and TC segmentation decoders only use signals from the two most relevant modalities to prevent noise from overwhelming the signal.

Without loss of generality, consider the current module to be the decoder at level $i$ (where $i$=1,2,3). The input data to the decoders for WT and TC segmentation includes four feature vectors:

(1) The common feature at level 4−$i$, denoted as $\mathbf{F}_{4-i}$,

(2) The fused feature from the previous level, $\mathbf{f}_{i-1}^{Y}$,

(3) The T2-channel feature at level 4−$i$, $\mathbf{F}_{4-i}^{T2}$,

(4) The other-channel feature at level 4−$i$, $\mathbf{F}_{4-i}^{X}$.

Let the output of the decoder be $\mathbf{f}_i^{Y}$. The decoders for the WT and TC segmentation can be represented by the following series of operations:

$$\varsigma_i^1 = \text{Conv}_1\left(\text{Concat}\left(\left(\mathbf{F}_{4-i}^{X} \otimes \mathbf{F}_{4-i}\right), \mathbf{F}_{4-i}\right)\right) \quad (i = 1,2,3). \quad (15)$$

$$\varsigma_i^2 = \text{Conv}_2\left(\text{Concat}\left(\left(\mathbf{F}_{4-i}^{T2} \otimes \mathbf{F}_{4-i}\right), \mathbf{F}_{4-i}\right)\right) \quad (i = 1,2,3). \quad (16)$$

$$\varsigma_i^3 = \text{Concat}\left(\text{Conv}_3\left(\text{Concat}(\varsigma_i^1, \varsigma_i^2)\right), \mathbf{F}_{4-i}^{X} \otimes \left(\mathbf{F}_{4-i}^{T2} \otimes \mathbf{F}_{4-i}\right)\right) (i = 1,2,3). \quad (17)$$

$$\mathbf{f}_i^Y = \text{Conv}_5\left(\text{Concat}(\mathbf{f}_{i-1}^Y, \text{Conv}_4(\boldsymbol{\varsigma}_i^3))\right) \qquad (i = 1,2,3). \qquad (18)$$

If $i > 1$, the fused feature from the previous level $\mathbf{f}_{i-1}^Y$ is the output of the previous decoder stage. If $i = 1$, for WT segmentation, $\mathbf{f}_{i-1}^Y$ is set to $\mathbf{F}_4^{WT}$ (as shown in Fig. 3), and for TC segmentation, $\mathbf{f}_{i-1}^Y$ is set to $\mathbf{F}_4^{TC}$ (also shown in Fig. 3).

For WT segmentation, the feature $\mathbf{F}_{4-i}^X$ corresponds to the FLAIR-channel feature at level $4-i$, denoted as $\mathbf{F}_{4-i}^{FLAIR}$. For TC segmentation, $\mathbf{F}_{4-i}^X$ corresponds to the T1Gd-channel feature at level $4-i$, denoted as $\mathbf{F}_{4-i}^{T1Gd}$. It is important to note that for the first decoder stage, the input channel features are from the third level.

**II. Decoder for ET segmentation**

The decoder for ET segmentation (see the "DecA" module in Fig. 3), has a structure similar to that of the WT and TC decoders, as depicted in Fig. 6(b). Based on the correlation analysis between ET segmentation and MRI modalities (see Fig. 2), the ET region is highly correlated with the T1Gd channel signal, while less correlated with other channels. Therefore, the decoder for ET segmentation only takes as input the common features and the T1Gd channel features to avoid interference from other channels.

Without loss of generality, we denote the current module as the decoder at level $i$ (where $i=1,2,3$). The input to the ET decoder includes three feature vectors:

(1) The common feature at level $4-i$, denoted as $\mathbf{X}_{4-i}$,

(2) The fused feature from the previous level, $\mathbf{f}_{i-1}^{ET}$,

(3) The T1Gd-channel feature at level $4-i$, $\mathbf{X}_{4-i}^{T1Gd}$.

Let the output of the decoder at level $i$ be denoted as $\mathbf{f}_i^Y$. The series of operations performed by the ET decoder can be expressed as follows:

$$\boldsymbol{\zeta}_i^1 = \text{Conv}_1\left(\text{Concat}\left((\mathbf{X}_{4-i}^{T1Gd} \otimes \mathbf{X}_{4-i}), \mathbf{X}_{4-i}\right)\right) \qquad (i = 1,2,3). \qquad (19)$$

$$\mathbf{f}_i^Y = \text{Conv}_2\left(\text{Concat}(\mathbf{f}_{i-1}^{ET}, \boldsymbol{\zeta}_i^1)\right) \qquad (i = 1,2,3). \qquad (20)$$

When $i = 1$ or $i = 2$, the inputs $\mathbf{X}_{4-i} = \mathbf{F}_{4-i}$, $\mathbf{X}_{4-i}^{T1Gd} = \mathbf{F}_{4-i}^{T1Gd}$. However, when $i = 3$, the input features $\mathbf{X}_{4-i}$ and $\mathbf{X}_{4-i}^{T1Gd}$ correspond to the edge-enhanced versions of features $\mathbf{F}_{4-i}$ and $\mathbf{F}_{4-i}^{T1Gd}$, respectively. If $i > 1$, the fused ET feature $\mathbf{f}_{i-1}^{ET}$ from the previous level is the output of the previous decoder stage. For $i = 1$, $\mathbf{f}_{i-1}^{ET}$ is set to $\mathbf{F}_4^{ET}$ (as shown in Fig. 3).

Among the three segmentation tasks, the performance for ET segmentation is the poorest. To improve ET segmentation accuracy, multiPI-TransBTS introduces a feature enhancement (FE) module into the ET encoder. This module is based on principles from information theory, where greater variability in feature values indicates a higher amount of information. While information entropy is a common measure of information, its computation involves logarithmic operations, which can be time-consuming. Therefore, we choose curvature as a computationally efficient alternative to measure information content.

The curvature is calculated using a convolution operation with a predefined kernel, as described in [56]. The convolution kernel is defined as:

$$\lambda = \frac{1}{16}\begin{bmatrix} -1 & 5 & -1 \\ 5 & -16 & 5 \\ -1 & 5 & -1 \end{bmatrix} \quad (21)$$

The feature enhancement module first calculates the curvature for each channel. It then selects the top *KC* channels with the highest curvature values, which are combined with the original feature maps and input into the decoder. This enhances the feature representation of channels containing the most significant information. In this study, we set *K*=0.5. The detailed process of the feature enhancement module is illustrated in Fig. 7.

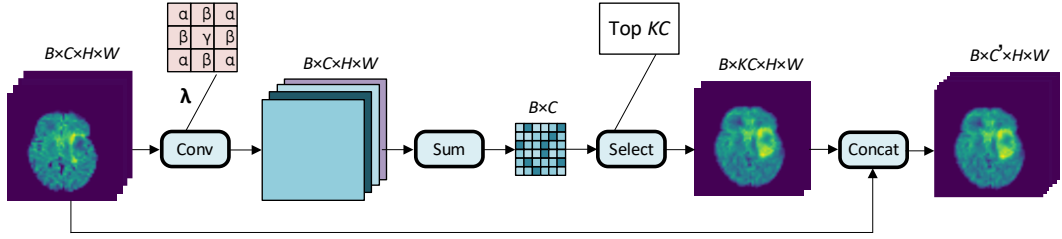

Fig. 7. Flow chart of feature enhancement module.

## 4.5 Loss function

Segmentation tasks can be framed as binary classification problems. The most commonly used loss function for classification is cross-entropy [57], which is rooted in the concept of information entropy introduced by Claude Shannon [58]. In the context of image segmentation, the target region is labeled as 1, and the background is labeled as 0. For each of the three tumor regions, we obtain a binary map comparing the multiPI-TransBTS predictions with the ground truth. To handle the issue where log(0) is undefined, the Binary Cross-Entropy (BCE) Loss with clipping for segmentation is expressed as [59]:

$$L_{\text{BCE}} = \sum_{i=1}^{N} -[y_i \cdot \max(\log(\hat{y}_i), -100) + (1 - y_i) \cdot \max(\log(1 - \hat{y}_i), -100)] \quad (22)$$

where $\hat{y}_i$ represents the segmentation result (0 or 1), $y_i$ is the ground truth label (0 or 1), $i$ denotes the pixel index, and $N$ is the total number of pixels.

Another classical loss function used for image segmentation model training is Dice Loss, which is based on the Dice coefficient [60]. For Boolean data, Dice Loss is mathematically defined as [61]:

$$L_{\text{Dice}} = 1 - \frac{2\sum_{i=1}^{N} y_i \hat{y}_i}{\sum_{i=1}^{N} y_i^2 + \sum_{i=1}^{N} \hat{y}_i^2} \quad (23)$$

where $\hat{y}_i$ represents the segmentation result (0 or 1), $y_i$ is the ground truth label (0 or 1), $i$ denotes the pixel index, and $N$ is the total number of pixels.

While BCE is well-suited for binary classification tasks, it can be dominated by the majority class (maybe non-target regions). On the other hand, Dice Loss effectively handles class imbalance but may have a less smooth gradient, making optimization more challenging compared to BCE Loss [41].

To leverage the advantages of both loss functions and mitigate their shortcomings, multiPI-TransBTS employs a combined loss function that integrates BCE Loss and Dice Loss:

$$L = \left(L_{\text{BCE}}^{\text{WT}} + L_{\text{Dice}}^{\text{WT}}\right) + \left(L_{\text{BCE}}^{\text{TC}} + L_{\text{Dice}}^{\text{TC}}\right) + \left(L_{\text{BCE}}^{\text{ET}} + L_{\text{Dice}}^{\text{ET}}\right). \tag{24}$$

# 5 EXPERIMENT SETUPS

## 5.1 Datasets

Similar to many previous studies [41, 62, 63], the training and testing datasets used in our experiments were obtained from the BraTS2019 and BraTS2020 datasets, which are part of the BraTS Challenge. These datasets consist of multi-institutional pre-operative MRI scans, primarily focused on the segmentation of intrinsically heterogeneous brain tumors in terms of appearance, shape, and histology.

All BraTS multimodal scans are provided in NIfTI format (.nii.gz) and include a) native T1-weighted (T1), b) post-contrast T1-weighted (T1Gd), c) T2-weighted (T2), and d) T2 Fluid Attenuated Inversion Recovery (FLAIR) volumes. Each dataset has been manually segmented by one to four raters following a standardized annotation protocol, and the annotations were subsequently validated by experienced neuro-radiologists. These annotations encompass the GD-enhancing tumor (ET — label 4), the peritumoral edema (ED — label 2), and the necrotic and non-enhancing tumor core (NCR/NET — label 1). The relationships between these labels and the final segmentation regions are outlined in Table 1.

BraTS2019 and BraTS2020 include 335 and 369 annotated subject samples, respectively. Consistent with prior research [62, 63], we randomly divided the provided brain tumor segmentation dataset into training, validation, and test sets using an 8:1:1 split ratio.

## 5.2 Evaluation metrics

Consistent with previous research [62-64], our evaluation employs three widely recognized metrics: Dice coefficient, Hausdorff distance (HD95), and Sensitivity.

The Dice coefficient evaluates the overall accuracy of the segmentation by measuring the overlap between the predicted segmentation and the ground truth. Mathematically, it is defined as:

$$\text{Dice} = \frac{1}{M} \sum_{M} \frac{2 \sum_{i=1}^{N} y_i \hat{y}_i}{\sum_{i=1}^{N} y_i^2 + \sum_{i=1}^{N} \hat{y}_i^2}, \tag{25}$$

where $\hat{y}$ represents the predicted segmentation result (0 or 1), and y is the ground truth label (0 or 1). Here, $i$ refers to the pixel index, $N$ is the total number of pixels per sample, and $M$ is the total number of samples.

The HD95 is the 95th percentile of the Hausdorff distance, which assesses how well the predicted boundaries align with the ground truth boundaries, focusing on worst-case discrepancies. Let $y \in Tr, \hat{y} \in Pr$, where $Pr$ and $Tr$ represent the predicted value and the true value of each point, respectively. The Hausdorff distance is given by:

$$\text{HD} = \max\left\{\max_{y \in Tr}\min_{\hat{y} \in Pr}\|y - \hat{y}\|, \max_{\hat{y} \in Pr}\min_{y \in Tr}\|y - \hat{y}\|\right\}. \qquad (26)$$

Finally, Sensitivity measures the model's ability to correctly identify positive regions, reflecting its capacity to detect all relevant features. It is defined as:

$$\text{Sensitivity} = \frac{TP}{TP + FN}, \qquad (27)$$

where $TP$, $FN$ represent the total number of true positives and false negatives, respectively.

# 6 EXPERIMENTAL RESULTS

The experiments were conducted on a server equipped with two NVIDIA Quadro RTX8000 (48GB) graphics cards. The code was implemented using PyTorch. The initial learning rate was set to 0.01, and the SGD optimizer was employed for learning rate optimization. A batch size of 12 was used during model training, and the total number of training epochs was set to 100.

## 6.1 Performance analysis

The performance comparison between multiPI-TransBTS and the baseline methods was conducted on two datasets. The baseline methods consist of 10 models, including both classical approaches, such as IVD-Net [65], UNet++ [66], and AttentionU [67], and recent representative models, like UCTransNet [68] and F$^2$Net [69]. The results for the BraTS2019 and BraTS2020 datasets are presented in Table 2 and Table 3, respectively. In the tables, bold red numbers indicate the best-performing model, while bold black numbers denote the second-best model. Note that

Table 2. Performance Comparison on the BraTS2019 Dataset

| Models | Dice | | | | HD95 | | | | Sensitivity | | | |
|---|---|---|---|---|---|---|---|---|---|---|---|---|
| | WT | TC | ET | Mean | WT | TC | ET | Mean | WT | TC | ET | Mean |
| UNet(15) [51] | 0.944 | 0.922 | 0.890 | 0.919 | 2.261 | 1.620 | 1.434 | 1.772 | 0.941 | 0.939 | 0.905 | 0.928 |
| IVD-Net (18) [65] | 0.928 | 0.902 | 0.870 | 0.900 | 2.340 | 1.670 | 1.430 | 1.810 | 0.917 | 0.923 | 0.882 | 0.907 |
| UNet++(19) [66] | 0.945 | 0.924 | 0.894 | 0.921 | 2.251 | 1.603 | 1.371 | 1.742 | 0.940 | 0.930 | 0.904 | 0.925 |
| AttentionU(19) [67] | 0.947 | 0.926 | 0.896 | 0.923 | **2.220** | 1.580 | **1.333** | 1.711 | 0.940 | 0.931 | 0.898 | 0.923 |
| TransUnet(21) [39] | 0.947 | 0.932 | 0.896 | 0.925 | 2.261 | 1.580 | 1.382 | 1.741 | 0.951 | 0.939 | **0.909** | 0.933 |
| UCTransNet (23) [68] | 0.949 | 0.934 | 0.903 | 0.929 | 2.209 | **1.558** | **1.329** | **1.699** | 0.944 | 0.938 | **0.909** | 0.930 |
| F$^2$Net(23) [69] | **0.952** | **0.941** | **0.900** | **0.931** | 2.222 | 1.652 | 1.348 | 1.741 | 0.954 | **0.949** | 0.905 | **0.936** |
| SPA-Net(24) [64] | 0.896 | 0.833 | 0.771 | 0.833 | 5.950 | 6.120 | 3.530 | 5.200 | - | - | - | - |
| EA-DFFTU(24) [62] | 0.902 | 0.815 | 0.737 | 0.818 | 4.900 | 4.600 | 5.400 | 4.967 | **0.955** | 0.814 | 0.713 | 0.827 |
| S$^2$CA-Net(24) [63] | 0.910 | 0.844 | 0.801 | 0.852 | 4.001 | 5.860 | 3.177 | 4.346 | - | - | - | - |
| Ours | **0.953** | **0.944** | **0.904** | **0.934** | **2.177** | **1.538** | 1.334 | **1.683** | **0.962** | **0.954** | 0.899 | **0.938** |

Note that higher values indicate better performance for Dice and Sensitivity, whereas lower values are better for HD95.

Table 3. Performance Comparison on the BraTS2020 Dataset

| Models | Dice | | | | HD95 | | | | Sensitivity | | | |
|---|---|---|---|---|---|---|---|---|---|---|---|---|
| | WT | TC | ET | Mean | WT | TC | ET | Mean | WT | TC | ET | Mean |
| UNet(15) [51] | 0.943 | 0.922 | 0.896 | 0.920 | 2.735 | 1.970 | 2.30 | 2.348 | 0.941 | 0.932 | 0.906 | 0.926 |
| IVD-Net (18) [65] | 0.931 | 0.906 | 0.874 | 0.904 | 2.771 | 2.527 | 2.153 | 2.484 | 0.920 | 0.926 | 0.885 | 0.910 |
| UNet++(19) [66] | 0.945 | 0.923 | 0.897 | 0.922 | 2.466 | 2.165 | 2.287 | 2.306 | 0.943 | 0.931 | 0.904 | 0.926 |
| AttentionU(19) [67] | 0.942 | 0.914 | 0.891 | 0.916 | 2.200 | **1.534** | 1.332 | 1.687 | 0.938 | 0.932 | 0.905 | 0.925 |
| TransUnet(21) [39] | 0.947 | 0.932 | 0.899 | 0.926 | 2.222 | 1.551 | 1.380 | 1.718 | 0.945 | 0.940 | 0.902 | 0.929 |
| UCTransNet (23) [68] | 0.947 | 0.934 | **0.904** | 0.928 | 2.211 | 1.550 | 1.343 | 1.701 | 0.940 | 0.938 | **0.913** | 0.930 |
| F$^2$Net(23) [69] | **0.951** | **0.940** | 0.902 | **0.931** | 2.195 | 1.544 | **1.313** | **1.684** | **0.953** | **0.951** | **0.916** | **0.940** |
| SPA-Net(24) [64] | 0.900 | 0.832 | 0.778 | 0.837 | 4.600 | 6.550 | 32.20 | 14.45 | - | - | - | - |
| EA-DFFTU(24) [62] | 0.937 | 0.855 | 0.806 | 0.866 | 3.700 | 3.200 | 2.500 | 3.133 | 0.943 | 0.928 | 0.907 | 0.926 |
| S$^2$CA-Net(24) [63] | 0.925 | 0.889 | 0.824 | 0.879 | 3.043 | 3.877 | 2.712 | 3.211 | - | - | - | - |
| Ours | **0.954** | **0.947** | **0.909** | **0.937** | **2.167** | **1.476** | **1.271** | **1.638** | **0.961** | **0.951** | 0.909 | **0.940** |

Note that higher values indicate better performance for Dice and Sensitivity, whereas lower values are better for HD95.

higher values indicate better performance for Dice and Sensitivity, while lower values are better for HD95.

From Table 2, it can be observed that multiPI-TransBTS achieves the best Dice, Hausdorff Distance (HD95), and Sensitivity scores for the segmentation of WT and TC. For ET, it achieves the highest Dice score.

Similarly, from Table 3, multiPI-TransBTS demonstrates superior performance, achieving the best Dice, HD95, and Sensitivity scores for the WT and TC segmentation. For ET segmentation, it again attains the highest Dice score and the best HD95.

Overall, multiPI-TransBTS consistently outperforms the baseline methods across the segmentation of three tumor regions. Although its performance varies slightly between the BraTS2019 and BraTS2020 datasets, the conclusions remain the same: multiPI-TransBTS achieves the best Dice, HD95, and Sensitivity scores for the WT and TC regions. For the ET region, it achieves the best Dice score but does not attain the highest Sensitivity, suggesting a strong ability to accurately segment the WT and TC regions but a limitation in fully capturing all relevant positive ET areas.

## 6.2 visualization of segmentation results

To provide an intuitive comparison of the performance of multiPI-TransBTS with other methods, we visualize the segmentation results on the BraTS2019 and BraTS2020 datasets. Due to space constraints, we randomly selected three samples from each dataset. In the figures, the blue, brown, and yellow regions represent necrotic tumor cores, enhancing tumors, and edematous regions, respectively. Red arrows indicate areas where our results outperform the baselines, while white arrows highlight some false positive results from the baseline methods.

The visualization of segmentation results on the BraTS2019 dataset is presented in Fig. 8, with the randomly selected samples being TCIA09_255, TCIA03_375, and TCIA08_406. As shown in Fig. 8, the segmentation results from multiPI-TransBTS are notably closer to the ground truth compared to the baselines.

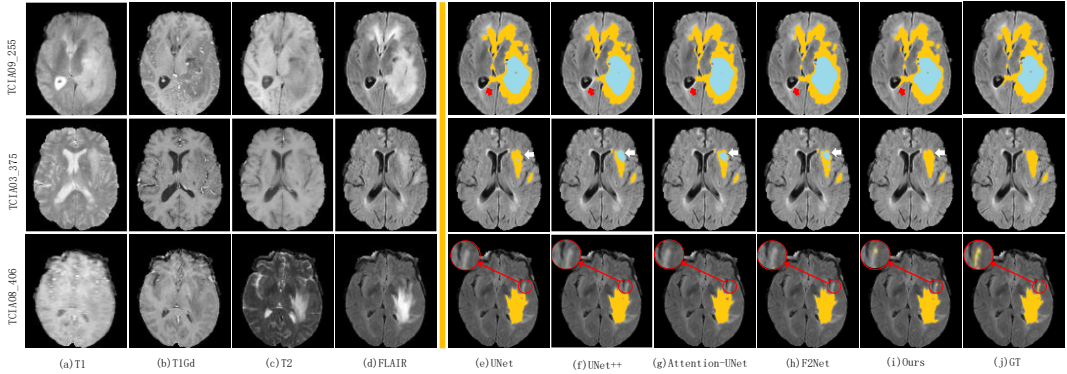

Fig. 8. Visualization of segmentation results on the BraTS2019 dataset. 'GT' in the figure refers to the ground truth.

Similarly, the visualization of segmentation results on the BraTS2020 dataset is presented in Fig. 9, with the randomly selected samples being 250, 342, and 345. The conclusions drawn from Fig. 9 are consistent with those from Fig. 8, further demonstrating the superior performance of multiPI-TransBTS.

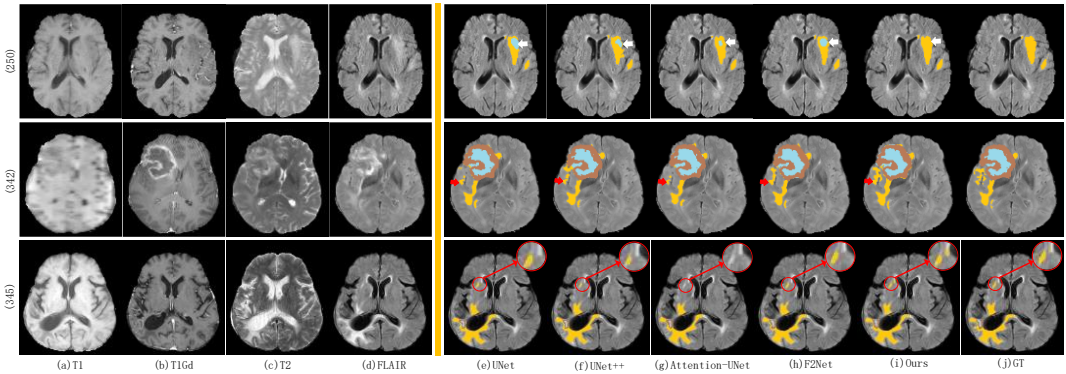

Fig. 9. Visualization of segmentation results on the BraTS2020 dataset. 'GT' in the figure refers to the ground truth.

## 6.3 Ablation study

To evaluate the contribution of each component within the multiPI-TransBTS framework, we conducted an ablation study. For this purpose, we developed three derived versions of the multiPI-TransBTS model: mT-AFF, mT-PCI, and mT-FE. The specific modifications for each version are as follows: (1) mT-AFF omits the Adaptive Feature Fusion (AFF) module; (2) mT-PCI excludes the Task-Specific Feature Introduction (TSFI) strategy; and (3) mT-FE removes the Feature Enhancement (FE) module. The results of these models on the BraTS2019 and BraTS2020 datasets are shown in Tables 4 and 5, respectively.

Tables 4 and 5 provide a detailed performance comparison between the original multiPI-TransBTS model and its derived versions. It is clear that the original multiPI-TransBTS generally

Table 4. Ablation study results on BraTS2019 dataset

| Models | Dice | | | | HD95 | | | | Sensitivity | | | |
|---|---|---|---|---|---|---|---|---|---|---|---|---|
| | WT | TC | ET | Mean | WT | TC | ET | Mean | WT | TC | ET | Mean |
| mT-AFF | **0.953** | 0.943 | **0.902** | 0.933 | 2.197 | 1.572 | 1.361 | 1.710 | **0.954** | 0.947 | **0.911** | 0.937 |
| mT-TSFI | 0.770 | 0.885 | 0.878 | 0.844 | 3.247 | 2.129 | 1.524 | 2.300 | 0.632 | **0.951** | 0.900 | 0.828 |
| mT-FE | **0.954** | **0.944** | **0.902** | **0.933** | 2.182 | **1.521** | **1.325** | **1.676** | 0.952 | 0.945 | 0.897 | 0.931 |
| multiPI-TransBTS | **0.953** | **0.944** | **0.904** | **0.934** | **2.177** | 1.538 | 1.334 | 1.683 | **0.962** | **0.954** | 0.899 | **0.938** |

Note that higher values indicate better performance for Dice and Sensitivity, whereas lower values are better for HD95.

Table 5. Ablation study results on BraTS2020 dataset

| Models | Dice | | | | HD95 | | | | Sensitivity | | | |
|---|---|---|---|---|---|---|---|---|---|---|---|---|
| | WT | TC | ET | Mean | WT | TC | ET | Mean | WT | TC | ET | Mean |
| mT-AFF | 0.950 | 0.941 | 0.904 | 0.933 | 2.177 | **1.538** | 1.315 | 1.676 | 0.951 | 0.944 | 0.907 | 0.934 |
| mT-TSFI | 0.861 | 0.805 | 0.875 | 0.847 | 2.879 | 2.815 | 1.640 | 2.444 | 0.773 | **0.965** | **0.913** | 0.884 |
| mT-FE | **0.953** | **0.946** | **0.907** | **0.935** | 2.169 | 1.539 | **1.311** | **1.673** | **0.960** | **0.952** | 0.896 | **0.936** |
| multiPI-TransBTS | **0.954** | **0.947** | **0.909** | **0.937** | **2.167** | **1.476** | **1.271** | **1.638** | **0.961** | 0.951 | **0.909** | **0.940** |

Note that higher values indicate better performance for Dice and Sensitivity, whereas lower values are better for HD95.

outperforms its variants, underscoring the importance and effectiveness of each integrated component within the framework.

# CONCLUSION

In this study, we introduced a novel framework for the BraTS challenge, multiPI-TransBTS: a Transformer-based architecture that integrates multi-physical information. This information includes general image attributes such as spatial and semantic information, along with specific information relevant to BraTS, particularly multimodal imaging knowledge.

To incorporate multi-physical information, multiPI-TransBTS leverages spatial-reduction attention modules, which enable the encoder to integrate multimodal information across different input sequences efficiently. Additionally, an adaptive feature fusion module uses both channel-wise and element-wise attention mechanisms to fuse complementary information from multiple modalities into a unified representation. The decoder employs a personality feature introduction strategy, which blends common features shared across tasks with individual features specific to each task. This process ensures that the abstracted features are optimally reconstructed into segmentation outputs.

Through rigorous experimental evaluations on two publicly available datasets, multiPI-TransBTS demonstrates superior performance compared to the state-of-the-art models. It achieves the best Dice, HD95, and sensitivity scores for WT and TC segmentation. For ET segmentation, it attains the best Dice and HD95 scores but does not achieve the highest sensitivity, indicating a trade-off between precision and recall. This trade-off presents a future research direction in balancing segmentation performance.

Overall, multiPI-TransBTS significantly outperforms existing models, validating the effectiveness of introducing multi-physical information into a Transformer-based framework for BraTS segmentation tasks. These advancements hold the potential to substantially enhance clinical outcomes for patients with brain tumors.

# ACKNOWLEDGMENTS

This work was supported by the National Social Science Foundation of China (20BXW097) and the Science and Technology Project Foundation of Chongqing Municipal Education Committee of China (KJZD-M202400601).